\documentclass[floatfix,twocolumn,prl,amsmath]{revtex4-2}
\usepackage{graphicx}
\usepackage{bm}
\usepackage{amssymb}
\usepackage{dcolumn}
\usepackage{subfigure}
\usepackage{multirow}
\usepackage{mathrsfs}
\DeclareMathAlphabet{\mathscrbf}{OMS}{mdugm}{b}{n}
\usepackage{color}

\usepackage{threeparttable}
\begin{document}
\newcommand{\intqspa}{\int\!\!\frac{\rmd^d q}{(2\pi)^d}}
\newcommand{\intqspathr}{\int\!\!\frac{\rmd^3 q}{(2\pi)^3}}
\newcommand{\intqspatwo}{\int\!\!\frac{\rmd^2 q}{(2\pi)^2}}
\newcommand{\intkspatwo}{\int\!\!\frac{\rmd^2 k}{(2\pi)^2}}
\newcommand{\intkspa}{\int\!\!\frac{\rmd^d k}{(2\pi)^d}}
\newcommand{\intkspapri}{\int\!\!\frac{\rmd^d k'}{(2\pi)^d}}
\newcommand{\vn}[1]{{\boldsymbol{#1}}}
\newcommand{\vht}[1]{{\boldsymbol{#1}}}
\newcommand{\matn}[1]{{\bf{#1}}}
\newcommand{\matnht}[1]{{\boldsymbol{#1}}}
\newcommand{\bege}{\begin{equation}}
\newcommand{\gretke}{G_{\vn{k} }^{\rm R}(\mathcal{E})}
\newcommand{\gret}{G^{\rm R}}
\newcommand{\gadv}{G^{\rm A}}
\newcommand{\gmat}{G^{\rm M}}
\newcommand{\gles}{G^{<}}
\newcommand{\ghat}{\hat{G}}
\newcommand{\sigmahat}{\hat{\Sigma}}
\newcommand{\glesone}{G^{<,{\rm I}}}
\newcommand{\glestwo}{G^{<,{\rm II}}}
\newcommand{\gspec}{G^{\rm S}}
\newcommand{\glesthree}{G^{<,{\rm III}}}
\newcommand{\magdir}{\hat{\vn{n}}}
\newcommand{\sigmaret}{\Sigma^{\rm R}}
\newcommand{\sigmales}{\Sigma^{<}}
\newcommand{\sigmalesone}{\Sigma^{<,{\rm I}}}
\newcommand{\sigmalestwo}{\Sigma^{<,{\rm II}}}
\newcommand{\sigmalesthree}{\Sigma^{<,{\rm III}}}
\newcommand{\sigmaadv}{\Sigma^{A}}
\newcommand{\ee}{\end{equation}}
\newcommand{\bal}{\begin{aligned}}
\newcommand{\defbar}{\overline}
\newcommand{\SM}{\scriptstyle}
\newcommand{\rmd}{{\rm d}}
\newcommand{\rme}{{\rm e}}
\newcommand{\eal}{\end{aligned}}
\newcommand{\crea}[1]{{c_{#1}^{\dagger}}}
\newcommand{\annihi}[1]{{c_{#1}^{\phantom{\dagger}}}}
\newcommand{\udot}{\overset{.}{u}}
\newcommand{\exponential}[1]{{\exp(#1)}}
\newcommand{\phandot}[1]{\overset{\phantom{.}}{#1}}
\newcommand{\phandag}{\phantom{\dagger}}
\newcommand{\Trace}{\text{Tr}}
\setcounter{secnumdepth}{2}
\title{Moment potentials for spectral density functional theory: Exploiting the momentum distribution of the uniform electron gas}
\author{Frank Freimuth$^{1,2}$}
\email[Corresp.~author:~]{f.freimuth@fz-juelich.de}
\author{Stefan Bl\"ugel$^{1}$}
\author{Yuriy Mokrousov$^{1,2}$}
\affiliation{$^1$Peter Gr\"unberg Institut and Institute for Advanced Simulation,
Forschungszentrum J\"ulich and JARA, 52425 J\"ulich, Germany}
\affiliation{$^2$ Institute of Physics, Johannes Gutenberg University Mainz, 55099 Mainz, Germany
}
\begin{abstract}
  In standard Kohn-Sham (KS) density-functional theory (DFT) the valence band satellites
  in Ni and Pd are missing, the band widths in Ni and Na are too large, and the
  formation of lower and upper Hubbard bands in SrVO$_3$ is not described.
  These spectral features may be corrected by constructing the spectral function from the
  first four spectral moments, which may be obtained numerically efficiently
  within a \textit{moment-functional based
  spectral density functional theory}
  (MFbSDFT).
  In order to obtain a suitable potential for the second moment, one may use
  existing models of the uniform electron gas (UEG). However, models for the
  third moment of the UEG are not yet available.
  Therefore, we show that reproducing the momentum distribution of the UEG
  within the two-pole approximation of the spectral function determines the
  third moment, when the second moment is given. This allows us to find a model
  for the second and third moment potentials, which reproduces the experimental
  spectra of Ni, Pd, Na, and SrVO$_3$, and which is consistent with the second moment of the
  UEG at low density.
  Additionally, we describe an efficient algorithm to
  compute the spectral function from the first $2P$ spectral moment matrices, which paves
  the way to increasing the accuracy of MFbSDFT by using more than 4 spectral moments.
\end{abstract}
\maketitle
\section{Introduction}

If the first $2P$ spectral moment matrices ($P=1,2,3,\dots$) are given,
one may calculate a spectral function with $PN$ poles that reproduces these
spectral moments~\cite{momentis}. Here, $N$ is the number of rows of the square
spectral moment matrices. The case $P=1$ corresponds to the KS-DFT without
correlation, i.e., only with the first-order exchange (the
so-called $X\alpha$
method~\cite{Connolly1977} with $\alpha=2/3$ when the exact exchange is replaced
by its local approximation~\cite{PhysRev.140.A1133}).
The cases $P\ge 2$ involve higher-order correlation functions
such as $\langle \Psi^{\dagger}_{\sigma}(\vn{r})\Psi_{\sigma'}^{\dagger}(\vn{r}')\Psi_{\sigma'}(\vn{r}'') \Psi_{\sigma}(\vn{r}''')  \rangle$~\cite{nickel_PhysRevB.40.5015,PhysRevB.69.045113,dynamical_corr_fun_farid}.

Several strategies exist to deal with these higher-order correlation functions:
First, one may use results from Monte-Carlo simulations of the uniform electron
gas (UEG) in order to evaluate some of the correlation functions and use
the single Slater Determinant approximation to estimate the remaining
terms~\cite{PhysRevB.69.045113}. Second, one may in some cases express the
higher-order correlation function in terms of the single-particle
spectral function~\cite{book_Nolting}.
Third, one may compute the higher-order correlation functions
self-consistently from suitably
defined higher-order spectral functions, which depend themselves on these
higher-order correlation functions~\cite{spectral}.
Fourth, one may argue that the higher spectral moments may be expressed in terms
of universal functionals of the spin density, which circumvents the
necessity to evaluate the higher-order correlation functions if these
universal functionals are known~\cite{momentis}.

The fourth strategy makes the various
techniques~\cite{PhysRevLett.77.3865,dur36734,PhysRevLett.129.246401,invdft_methods}
that have been developed
in order to approach the universal exchange correlation potential of
KS-DFT and its
time-dependent generalizations
available to find suitable potentials, from which the spectral
moments may be obtained.
Within this method, the simplest possible expression for the moment matrices
is~\cite{momentis}
\bege
\vn{M}^{(I)}=\left[ \vn{M}^{(1)} \right]^{I}+\vn{M}^{(I+)},
\ee
where $\vn{M}^{(1)}$ is the first spectral moment matrix, $I=2,3,\dots$,
\bege
M^{(I+)}_{nm}=\int d^3 r \mathcal{V}^{(I+)}(\vn{r})
\left[\phi_{n}(\vn{r})\right]^{*}
\phi_{m}(\vn{r}),
\ee
$\mathcal{V}^{(I+)}(\vn{r})$ are the moment potentials, and $\phi_{n}(\vn{r})$
are orthonormalized basis functions.
The first spectral moment matrix, $\vn{M}^{(1)}$, is simply the
standard KS-Hamiltonian with first order exchange only, i.e., without
correlation. Within the local density approximation, the moment
potentials $\mathcal{V}^{(I+)}(\vn{r})$ depend on position $\vn{r}$ only through the
local density $n(\vn{r})$: $\mathcal{V}^{(I+)}(\vn{r})=\mathcal{V}^{(I+)}(n(\vn{r}))$.
Possible extensions are the inclusion of density gradients in the local
potentials $\mathcal{V}^{(I+)}(\vn{r})$ and the use of non-local potentials
$\mathcal{V}^{(I+)}(\vn{r},\vn{r}')$.

Assuming that the potentials $\mathcal{V}^{(I+)}(\vn{r},\vn{r}')$
(or $\mathcal{V}^{(I+)}(\vn{r})$ in the local approximation)
are universal functionals of the charge density, one may obtain them from
the UEG.
However, in the UEG only
$\vn{M}^{(1)}$ and
$\vn{M}^{(2)}$
have been studied in detail so far~\cite{PhysRevB.69.045113}.
Nevertheless, one may vary suitable trial functions $\mathcal{V}^{(I+)}(\vn{r})$
until spectral properties known from experiment are reproduced. If MFbSDFT is used in this way,
it is a similar approach like LDA+$U$, where the appropriate $U$ is often
determined by trying rather than by calculation. Using this trial approach,
we have reproduced the experimental band-width and exchange splitting in Ni, as well as
the position of the valence band satellite peaks~\cite{momentis}.

Finding the moment potentials $\mathcal{V}^{(I+)}(\vn{r},\vn{r}')$ for a given spectrum
in this way
may be considered as a generalization of the inverse problem of DFT~\cite{invdft_methods}.
Thus, the trial approach to MFbSDFT may also be considered as a form of \textit{inverse} MFbSDFT.
However, since we need a moment potential for every $I$ ($I=2,3,\dots$), the functional
space of the trial potentials $\mathcal{V}^{(I+)}(\vn{r},\vn{r}')$ is now much larger:
If we use the first four spectral moment matrices, we need to find two potentials,
$\mathcal{V}^{(2+)}(\vn{r},\vn{r}')$
and $\mathcal{V}^{(3+)}(\vn{r},\vn{r}')$, simultaneously.

In this work we show that the simultaneous search for suitable $\mathcal{V}^{(2+)}(\vn{r})$
and $\mathcal{V}^{(3+)}(\vn{r})$ may be simplified by requiring $\mathcal{V}^{(3+)}(\vn{r})$
to reproduce the known momentum
distribution $n_{k}$ of the interacting UEG for a given $\mathcal{V}^{(2+)}(\vn{r})$.
Consequently, only $\mathcal{V}^{(2+)}(\vn{r})$ needs to be varied in this simplified
version of \textit{inverse} MFbSDFT, because $\mathcal{V}^{(3+)}(\vn{r})$ is determined
by $n_{k}$ when $\mathcal{V}^{(2+)}(\vn{r})$ is given.
The momentum distribution of the UEG is known from Monte Carlo
simulations~\cite{PhysRevB.50.1391,PhysRevB.56.9970,PhysRevLett.107.110402}
and
suitable analytical models have been developed for it~\cite{PhysRevB.66.235116}.

Compared with the many established techniques for the treatment of correlations
in \textit{ab-initio} theory, we expect the MFbSDFT approach to have
an advantage when the system size is large due to its small
computational cost~\cite{momentis}. While the LDA+$U$ approach is also
well suited to study large systems, it corrects mostly the Fermi surface e.g.\
in Ni~\cite{PhysRevLett.87.216405}, but not the bandwidth.
While response functions such as the AHE can often be computed in
good agreement with experiment if the
Fermi surface is described well~\cite{PhysRevB.84.144427},
the correct bandwidths are certainly important for
laser-excited responses such
as photocurrents~\cite{PhysRevB.104.L220405}
and torques~\cite{PhysRevB.103.174429}.

Promising applications of the MFbSDFT approach may therefore be expected
e.g.\ in the field of antiferromagnetic spintronics~\cite{Jungwirth2016}.
Here, correlation effects are often strong, because the suppressed hybridization
between nearest-neigbor magnetic lattice sites reduces the bandwidth when compared
with the corresponding ferromagnetic state.
While ground state properties such as crystal structure, lattice parameters, and
magnetism are sometimes well described by GGA in antiferromagnets
such as FeRh~\cite{PhysRevB.94.014109}, the proper description
of the correlation effects in the spectra
is expected to be important for the calculation of response functions.
At the same time the \textit{ab-initio} description of these response functions
for realistic spintronic devices requires often large unit cells.
For example the device structure
FeRh(10~nm)/Cu(2~nm)/Ni$_{80}$Fe$_{20}$(10~nm)/Al(3~nm)~\cite{PhysRevApplied.18.024075} 
would require a forbidding amount of computational time in several
beyond-DFT methods~\cite{beyond_dft}.

The rest of this paper is structured as follows.
In Sec.~\ref{sec_mfbsdft} we briefly review the MFbSDFT approach.
In Sec.~\ref{sec_ueg_first_two} we review existing models of the
first two spectral moments of the UEG.
In Sec.~\ref{sec_ueg_twopole} we show how to determine the
third moment within the two-pole approximation when the momentum
distribution and the second moment are given.
In Sec.~\ref{sec_constru_pot} we discuss various suitable model potentials.
In Sec.~\ref{sec_selfcons_corr} we show how to obtain some higher-order
correlation functions from the spectral function in the UEG.
In Sec.~\ref{sec_algo_general} we present an efficient algorithm to
construct the spectral function from the first $2P$ spectral moment matrices.
In Sec.~\ref{sec_results} we discuss our results for Na, SrVO$_3$, Ni, and
Pd.
After discussing conclusions and open questions in
Sec.~\ref{sec_discussion},
this paper ends with a summary in Sec.~\ref{sec_summary}.

\section{Theory}
\subsection{MFbSDFT}
\label{sec_mfbsdft}
In this section, we briefly review the
key components of the MFbSDFT approach, which
are given in detail in Ref.~\cite{momentis}.
We assume that the spectral
moments are expressed in a basis set of $N$ orthonormalized
functions $\phi_{\vn{k}n}(\vn{r})$ and
thus given by
$N\times N$ matrices. Therefore, the zeroth moment $\vn{M}_{\vn{k}}^{(0)}$
is simply the unit matrix.
In order to compute the spectral function
approximately
from the first four spectral moments ($\vn{M}_{\vn{k}}^{(0)}$, $\vn{M}_{\vn{k}}^{(1)}$, $\vn{M}_{\vn{k}}^{(2)}$, and $\vn{M}_{\vn{k}}^{(3)}$),
we diagonalize 
the hermitean
$2N\times 2N$ matrix~\cite{momentis}
\bege\label{eq_2n2n_mat}
\vn{\mathcal{H}}_{\vn{k}}=\left(
\begin{array}{cc}
\vn{M}_{\vn{k}}^{(1)}
&\vn{B}_{1\vn{k}} \\
\vn{B}_{1\vn{k}}^{\dagger} &\vn{D}_{1\vn{k}}
\end{array}
\right),
\ee
where 
$\vn{B}_{1\vn{k}}=\vn{\mathcal{U}}_{\vn{k}}\sqrt{\vn{\mathcal{D}}_{\vn{k}}}$,
$\vn{B}_{2\vn{k}}=[\vn{M}_{\vn{k}}^{(3)}-\vn{M}_{\vn{k}}^{(2)}\vn{M}_{\vn{k}}^{(1)}][\vn{B}_{1\vn{k}}^{\dagger}]^{-1}$,
and
$\vn{D}_{1\vn{k}}=\vn{B}_{1\vn{k}}^{-1}[\vn{B}_{2\vn{k}}-\vn{M}_{\vn{k}}^{(1)}\vn{B}_{1\vn{k}}]$.
Here, $\vn{\mathcal{D}}_{\vn{k}}$ is a diagonal matrix,
and $\vn{\mathcal{U}}_{\vn{k}}$ is a unitary matrix
so that $\vn{\mathcal{U}}_{\vn{k}}\vn{\mathcal{D}}_{\vn{k}}\vn{\mathcal{U}}_{\vn{k}}^{\dagger}=\vn{M}_{\vn{k}}^{(2)}-\vn{M}_{\vn{k}}^{(1)}\vn{M}_{\vn{k}}^{(1)}$.
We suppress the spin index in $\vn{\mathcal{H}}_{\vn{k}}$
and in the moments $\vn{M}_{\vn{k}}^{(I)}$
for notational brevity.

We may write $\vn{\mathcal{H}}_{\vn{k}}=\vn{U}\vn{D}\vn{U}^{\dagger}$,
where $\vn{D}$ is a diagonal matrix with the $i$-th eigenvalue of $\vn{\mathcal{H}}_{\vn{k}}$
in the $i$-th row, i.e., $D_{ij}=\delta_{ij}E_{j}$.
The eigenenergies of $\vn{\mathcal{H}}_{\vn{k}}$
determine the poles of the
single-particle spectral function,
which is given by
\bege\label{eq_final_spec_general}
\frac{S_{ij}(E)}{\hbar}=\sum_{l=1}^{2N}a_{l}\mathcal{V}_{il}\mathcal{V}^{*}_{jl}\delta(E-E_{l}).
\ee
Here,
\bege\label{eq_statevecs_general}
\mathcal{V}_{i j}=\frac{U_{i j}}{\sqrt{a_{j}}},
\ee
with $i=1,...,N$, $j=1,...,2N$,
and
\bege\label{eq_spec_wei}
a_{j}=\sum_{i=1}^{N} U_{i j} [U_{i j}]^{*}
\ee
are
the spectral weights.
The spectral density is given by
\bege
A(E)=\frac{1}{\mathcal{N}}
\sum_{\vn{k}}A_{\vn{k}}(E)
\ee
where $\mathcal{N}$ is the number of $\vn{k}$ points,
\bege
A_{\vn{k}}(E)=
  \sum_{l=1}^{2N}a_{l}\delta(E-E_l),
\ee
and we suppress the $\vn{k}$-index in $a_{l}$ and $E_l$ for notational
convenience.

\subsection{The first two moments in the homogeneous electron gas}
\label{sec_ueg_first_two}
Expressions for the spectral moments in the homogeneous electron
gas have been given in Refs.~\cite{PhysRevB.69.045113,dynamical_corr_fun_farid}.
Ref.~\cite{PhysRevB.69.045113} expresses the
contributions to the second moment in terms of the
exchange self-energy, the
pair correlation function, and a remaining higher correlation function,
which has been treated within the single-Slater-determinant approximation.

Explicitly, we obtain $M_{k}^{(2)}=\left[M_{k}^{(1)}\right]^{2}+M_{k}^{(2+)}$,
where
\bege\label{eq_m2p_ueg}
M^{(2+)}_{k}=\Sigma^{(1)}_{\rm loc}+\Sigma^{(1)}_{{\rm nl},k}
\ee
comprises a $k$-independent contribution~\cite{PhysRevB.69.045113}
\bege\label{eq_sigmaloc}
\Sigma^{(1)}_{\rm loc}=
\left[
\frac{\hbar^2}{2 m a^2_{\rm B}}
\right]^2
\frac{32}{3\pi^2}\frac{k_{\rm F}}{(\bar{\alpha} r_s)^2}
\int_{0}^{\infty}\frac{d\,k'}{k'^2}S(k')
\ee
and a $k$-dependent contribution~\cite{PhysRevB.69.045113}
\bege\label{eq_sigmanlk}
\begin{aligned}
  &\Sigma^{(1)}_{{\rm nl},k}=
-\left[
\Sigma_{k}^{(0)}
\right]^2
-
\left[
\frac{\hbar^2}{2 m a_{\rm B}^2}
\right]^2
\frac{2}{(\bar{\alpha} r_{s})^3 \pi k k_{\rm F}}\times\\
&\times\int_{0}^{\infty}d\,k'\,k'\,\Sigma_{k'}^{(0)}(1-2n_{k'})
\ln
\left|
\frac{k-k'}{k+k'}
\right|.
\end{aligned}
\ee
Here, $\bar{\alpha}=[4/(9\pi)]^{1/3}$, $k_{\rm F}=(\bar{\alpha} r_s a_{\rm B})^{-1}$,
$S(k)$ is the structure factor, which is obtained from the
pair correlation function $g(r)$ by the Fourier transformation
\bege\label{eq_struc_fac}
S(k)=1+4\pi n\int d\,r r
\frac{\sin(kr)}{k}[g(r)-1],
\ee
and
\bege\label{eq_sigma0}
\Sigma_{k}^{(0)}=-
\frac{\hbar^2}{2 m a^2_{\rm B}}
\frac{2}{\pi k k_{\rm F}\bar{\alpha} r_s}
\int_{0}^{\infty}
d\,k'\,k' \,n_{k'}
\ln
\left|
\frac{k+k'}{k-k'}
\right|
\ee
is the exchange self-energy.

For the momentum distribution function $n_{k}$ occurring
in Eq.~\eqref{eq_sigmanlk} and in Eq.~\eqref{eq_sigma0}
several models are available.
The simplest possibility is the Hartree-Fock
model $n_{k}=\theta(k_{\rm F}-k)$, where $\theta(k)$ is the 
step function.
Monte Carlo simulations have been
used to compute $n_{k}$ for the
homogeneous electron gas~\cite{PhysRevB.50.1391}
showing that deviations of $n_{k}$ from the Hartree-Fock model
can be significant (see e.g.\ Fig.~\ref{fig_nk_ueg}).
A refined model for $n_{k}$,
which is close to the results of Monte Carlo simulations,
is based
on the Kulik function~\cite{PhysRevB.66.235116}.
Similarly, several models for the pair correlation function $g(r)$
in Eq.~\eqref{eq_struc_fac} are
available~\cite{PhysRevB.46.12947,PhysRevB.61.7353}.

$M_{k}^{(2+)}$ as given by Eq.~\eqref{eq_m2p_ueg}
is positive definite (see Fig.~7 in Ref.~\cite{PhysRevB.69.045113}).
This is important, because
the algorithm for the construction of the spectral function
from the first 4 spectral moment matrices that we have
developed in Ref.~\cite{momentis} requires $M^{(2+)}$ to
be positive definite.

The first moment of the homogeneous electron gas is given
by~\cite{PhysRevB.69.045113}
\bege
M_{k}^{(1)}=\frac{\hbar^2 k^2}{2m}+\Sigma_{k}^{(0)}.
\ee
Therefore, $\left[ M_{k}^{(1)}\right]^2$
contains the term $\left[\Sigma_{k}^{(0)}\right]^2$.
When we evaluate $M_{k}^{(2)}=\left[M_{k}^{(1)}\right]^{2}+M_{k}^{(2+)}$
this term $\left[\Sigma_{k}^{(0)}\right]^2$ cancels
with the term $-\left[\Sigma_{k}^{(0)}\right]^2$
in $\Sigma^{(1)}_{{\rm nl},k}$ (see Eq.~\eqref{eq_sigmanlk}).
When we construct the moment potentials for MFbSDFT, we should therefore
aim at a similar cancellation of the square of the exchange self energy
in the second moment.

In order to construct $\mathcal{V}^{(2+)}(\vn{r})$ from
Eq.~\eqref{eq_m2p_ueg}
we recall that in the local density approximation to DFT
the first order exchange potential is given by~\cite{PhysRev.140.A1133}
\bege\label{eq_vx_lda}
V_{x}(\vn{r})=\Sigma_{k=k_{\rm F}=(\bar{\alpha} a^{\phantom{B}}_{\rm B} r_s(\vn{r}))^{-1}}^{(0)}
=\frac{\hbar^2}{ m a_{\rm B}^2}
\left[
\frac{3}{2\pi}
\right]^{\frac{2}{3}}
\frac{1}{r_s(\vn{r})},
\ee
i.e., Eq.~\eqref{eq_sigma0} is evaluated at $k=k_{\rm F}$ using the
Hartree-Fock model of the momentum distribution $n_{k}=\theta(k_{\rm F}-k)$.
Therefore, one possible way to obtain the moment potential
is
\bege\label{eq_vf2+}
\mathcal{V}^{(2+)}_{\rm F}(\vn{r})=\Sigma^{(1)}_{\rm loc}+\Sigma^{(1)}_{{\rm nl},k=k_{\rm F}=(\bar{\alpha} a^{\phantom{B}}_{\rm B} r_s(\vn{r}))^{-1}},
\ee
where the Hartree-Fock model $n_{k}=\theta(k_{\rm F}-k)$
is used to evaluate Eq.~\eqref{eq_sigmanlk}.

Since the difference between the moment distribution $n_{k}$
and the simple Hartree-Fock model $n_{k}=\theta(k_{\rm F}-k)$
may be significant, Ref.~\cite{PhysRevB.69.045113}
considered using the refined model for $n_{k}$ presented
in Ref.~\cite{PhysRevB.66.235116} in order to evaluate
the exchange self energy Eq.~\eqref{eq_sigma0} and
found that the exchange self energy may indeed be modified
by this refinement (see Fig.~3 in Ref.~\cite{PhysRevB.69.045113}).
However, the refined model of Ref.~\cite{PhysRevB.69.045113}
does not
impose $k_{\rm F}$ as a cutoff for $k$, i.e., the Fermi sea of the
correlated electron gas may contain wavenumbers larger than $k_{\rm F}$ by any amount.
Consequently, there does not seem to be
a good reason to believe that evaluating Eq.~\eqref{eq_sigma0}
at $k=k_{\rm F}$
with the refined model of $n_{k}$
gives the right answer for the refinement of the LDA exchange self energy.

Therefore, we recall the derivation of Eq.~\eqref{eq_vx_lda}
from the minimization of the total energy~\cite{PhysRev.140.A1133}:
\bege\label{eq_epsx_mini}
V_{x}(\vn{r})=\frac{\partial [n(\vn{r})\epsilon_x(n(\vn{r}))]}{\partial n(\vn{r})},
\ee
where $\epsilon_x$ is the exchange energy per particle, which
may be obtained from
\bege
\epsilon_{x}=\frac{\int d k' k'^2 n_{k'} \Sigma^{(0)}_{k'}}{2\int d k' k'^2 n_{k'}}
\ee
even when the refined model of $n_{k}$ is used.
Similarly, one may consider
to use
\bege\label{eq_va2+}
\mathcal{V}_{\rm A}^{(2+)}=\Sigma_{\rm loc}^{(1)}+
\frac{\partial\left[n(\vn{r})
  \frac{\int d k' k'^2 n_{k'} \Sigma^{(1)}_{{\rm nl},k'}}{2\int d k' k'^2 n_{k'}}\right]
  }{\partial n(\vn{r})}
\ee
to obtain the second moment potential.

In Table~\ref{tab_compare_ueg}
we compare $\mathcal{V}_{\rm F}^{(2+)}$
and $\mathcal{V}_{\rm A}^{(2+)}$
in the first two columns, where we use
the Hartree-Fock model for $n_k$ and
the Perdew-Wang model~\cite{PhysRevB.46.12947} for $g(r)$.
Even with the Hartree-Fock model $\mathcal{V}_{\rm F}^{(2+)}$
and $\mathcal{V}_{\rm A}^{(2+)}$
differ. This suggests that Eq.~\eqref{eq_va2+} is not the correct
way to construct a local density approximation for $M^{(2+)}$.
If the MFbSDFT approach may be combined with variational techniques has
not yet been explored, therefore it is an open question if a correct
version of Eq.~\eqref{eq_va2+} can be derived from a variational principle.
Consequently, we stick to $\mathcal{V}_{\rm F}^{(2+)}$ computed with the Hartree Fock model
of $n_k$.

\begin{threeparttable}
\caption{
  Comparison between the second moment potentials
  derived from the UEG
  and the square of the correlation potential $[V_{c}]^2$.
  All numbers are in units of Ry$^{2}$,
  where Ry=$\frac{\hbar^2}{2 m a_{\rm B}^2}$=13.6~eV.
}
\label{tab_compare_ueg}
\begin{ruledtabular}
\begin{tabular}{c|c|c|c|c|}
$r_{s}$
&$\mathcal{V}_{\rm F}^{(2+)}$
&$\mathcal{V}_{\rm A}^{(2+)}$
  &$15[V_{c}]^{2}$
  &$100[V_{c}]^{2}$
  \\
  \hline
   1 &4.56 &9.6 &0.28 & 1.84
\\
\hline
   2 &0.92 &2.2 &0.16 & 1.07
\\
\hline
   3 &0.35 &0.91 &0.11 &  0.74
\\
\hline
   4 &0.18 &0.49 &0.084 & 0.56
   \\
   \hline
   5 &0.104 &0.3 &0.067 & 0.45
   \\
\hline
   6 &0.067 &0.19 &0.055 & 0.37
   \\   
\hline
   7 &0.046 &0.14 &0.046 & 0.31
\\
\hline
   8 &0.033 &0.1 &0.04 & 0.26
\\
\hline
   9 &0.025 &0.079 &0.035 & 0.23
\\
\end{tabular}
\end{ruledtabular}
\end{threeparttable}

$\mathcal{V}_{\rm F}^{(2+)}$ is a parameter-free model for $\mathcal{V}^{(2+)}$.
In  Table~\ref{tab_compare_ueg} the square of the correlation potential
is shown in the last two columns for comparison.
The potential $100 [V_{c}]^2$ is close to $\mathcal{V}_{\rm F}^{(2+)}$
for density parameters around $r_{s}=2$.
In Ref.~\cite{momentis}
we have used $100 [V_{c}]^2$ for SrVO$_3$,
but $15 [V_{c}]^2$ for Ni.
At $r_s=7$ we find $\mathcal{V}_{\rm F}^{(2+)}\approx 15 [V_{c}]^2$.
Our different choice of $\mathcal{V}^{(2+)}$ for SrVO$_3$
and Ni in Ref.~\cite{momentis} is therefore
qualitatively consistent with the expectation that the strongly
correlated system SrVO$_3$ is described by a smaller value of $r_s$
than Ni. However, as will become clear in
section~\ref{sec_ueg_twopole}, the potentials $\mathcal{V}^{(2+)}$
and $\mathcal{V}^{(3+)}$ need to correspond to each other: An increase
of $\mathcal{V}^{(2+)}$ may sometimes be compensated by an increase of $\mathcal{V}^{(3+)}$.
Therefore, comparing only $\mathcal{V}^{(2+)}$ between two systems may be
insufficient when $\mathcal{V}^{(3+)}$ differs as well.

The potentials $\mathcal{V}_{\rm F}^{(2+)}$
and $\mathcal{V}_{\rm A}^{(2+)}$
both rise stronger than $r_s^{-2}$ as $r_s$ decreases,
while $V_{c}$ rises weaker than $r_s^{-2}$.
Therefore, at $r_s=7$ we find $\mathcal{V}_{\rm F}^{(2+)}\approx 15 [V_{c}]^2$,
while $\mathcal{V}_{\rm F}^{(2+)}> 15 [V_{c}]^2$ for $r_s<7$,
and  $\mathcal{V}_{\rm F}^{(2+)}< 15 [V_{c}]^2$ for $r_s>7$.
Similarly, we find $\mathcal{V}_{\rm F}^{(2+)}< 100 [V_{c}]^2$
for $r_s>1$, but $\mathcal{V}_{\rm F}^{(2+)}> 100 [V_{c}]^2$
for $r_s<1$.

In Ref.~\cite{momentis} we tried to reproduce the experimental
spectra of Ni and SrVO$_3$ with $\mathcal{V}^{(2+)}\propto r_s^{-2}$ and
with $\mathcal{V}^{(2+)}\propto [V_c]^2$. We found it significantly easier to
reproduce the experimental spectra with the model  $\mathcal{V}^{(2+)}\propto [V_c]^2$.
This is surprising, because $[V_c]^2$ rises slower than $r_s^{-2}$
as $r_s$ decreases, while
both $\mathcal{V}_{\rm F}^{(2+)}$
and $\mathcal{V}_{\rm A}^{(2+)}$
rise stronger than $r_s^{-2}$ as $r_s$ decreases.
A priori, the following explanations may contribute to this observation:
First, Ref.~\cite{PhysRevB.69.045113}
estimates the
term $\langle c^{\dagger}_{\vn{k}-\vn{q}'\sigma}\rho_{\vn{q}} c_{\vn{k}-\vn{q}-\vn{q}'\sigma}  \rangle$ based on the single-Slater-determinant approximation.
We will get back to this point in Sec.~\ref{sec_selfcons_corr}.
Second, we will show in Sec.~\ref{sec_ueg_twopole}
that $\mathcal{V}^{(2+)}$ and  $\mathcal{V}^{(3+)}$ should be constructed
consistently, and a strong rise in $\mathcal{V}^{(2+)}$ requires an
even stronger rise in $\mathcal{V}^{(3+)}$. Thus, it is possible that
the potentials $\mathcal{V}_{\rm F}^{(2+)}$
and $\mathcal{V}_{\rm A}^{(2+)}$
rise too strongly at small $r_s$ for a local density approach to be possible:
A local density approach is only meaningful if the inhomogeneities
in the potential remain below a certain bound. However, it could be
that the steeper potentials $\mathcal{V}_{\rm F}^{(2+)}$
and $\mathcal{V}_{\rm A}^{(2+)}$ require inhomogeneities in $\mathcal{V}^{(3+)}$
that are too large for a local density approach. In this case, it
could be that $\mathcal{V}_{\rm F}^{(2+)}$
and $\mathcal{V}_{\rm A}^{(2+)}$ can be used in a generalized approach that
makes use also of the density gradients to construct the moment potentials.
Third, it might be that $\mathcal{V}_{\rm F}^{(2+)}$
or $\mathcal{V}_{\rm A}^{(2+)}$ are close to the universal moment functional if
a sufficient number of moments is used: It could be that performing \textit{inverse}
MFbSDFT with only the first 4 moments will not yield the universal moment potential
but instead a potential which is different from it because the number of moments used
is limited to 4.

\subsection{The uniform electron gas in the two-pole approximation}
\label{sec_ueg_twopole}

Within the two-pole approximation the spectral function of the
UEG is given by
\bege\label{eq_spec_fun_ueg_twopole}
\frac{S_{k}(E-\mu)}{\hbar}=a_{k,1}\delta(E-E_{k,1})+a_{k,2}\delta(E-E_{k,2}),
\ee
where we assume $E_{k,1}<E_{k,2}$.
We consider the
UEG without spin polarization and therefore we leave away the spin index
for notational convenience.
The spectral weights $a_{k,i}$ and energies $E_{k,i}$
may be computed numerically by diagonalizing a 2$\times$2 matrix
as described in Sec.~\ref{sec_mfbsdft}.
Alternatively, they may also be obtained analytically:
\bege
E_{k,1}=\frac{1}{2}\frac{M_{k}^{(1)}M_{k}^{(2)}-M_{k}^{(3)}+L}{[M_{k}^{(1)}]^2-M_{k}^{(2)}}
\ee
\bege
E_{k,2}=\frac{1}{2}\frac{M_{k}^{(1)}M_{k}^{(2)}-M_{k}^{(3)}-L}{[M_{k}^{(1)}]^2-M_{k}^{(2)}}
\ee
\bege
\label{eq_ak1_analy}
\begin{aligned}
  a_{k,1}=\frac{2[M_{k}^{(1)}]^3-3M_{k}^{(1)}M_{k}^{(2)}+M_{k}^{(3)}+L}{2L}
\end{aligned}  
\ee
\bege
a_{k,2}=\frac{-2[M_{k}^{(1)}]^3+3M_{k}^{(1)}M_{k}^{(2)}-M_{k}^{(3)}+L}{2L}
\ee
where
\bege
\begin{aligned}
  L=\Bigl[& 4\left[M_{k}^{(1)}\right]^3 M_{k}^{(3)}-3\left[M_{k}^{(1)}\right]^2  \left[M_{k}^{(2)}\right]^2 \\
    &-6M_{k}^{(1)}M_{k}^{(2)}M_{k}^{(3)}+\left[M_{k}^{(3)}\right]^2+4\left[M_{k}^{(2)}\right]^3  \Bigr]^{\frac{1}{2}}.
  \end{aligned}
\ee

\begin{figure}
\includegraphics[angle=0,width=\linewidth]{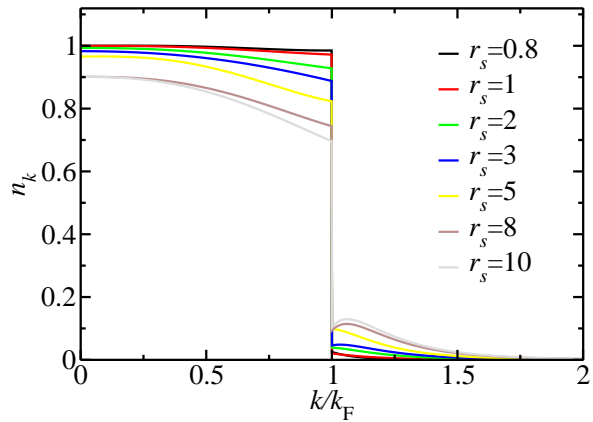}
\caption{\label{fig_nk_ueg}
  Momentum distribution function of the UEG according
  to the fits to the Monte Carlo data in Ref.~\cite{PhysRevB.56.9970}. 
}
\end{figure}

As discussed in Sec.~\ref{sec_ueg_first_two}
useful models for $M_{k}^{(1)}$ and $M_{k}^{(2)}$
have been developed for the UEG.
However, a model for $M_{k}^{(3)}$ is still missing.
Therefore, we consider the momentum distribution function
\bege\label{eq_nk_general}
n_k=\langle c^{\dagger}_{k} c_{k}   \rangle=\int dE f(E) S_{k}(E-\mu),
\ee
which has been investigated in detail for the UEG using
Monte Carlo methods.
Inserting the two-pole approximation Eq.~\eqref{eq_spec_fun_ueg_twopole}
into Eq.~\eqref{eq_nk_general}
we obtain
\bege\label{eq_nk_twopole}
n_k=f(E_{k,1})a_{k,1}+f(E_{k,2})a_{k,2}.
\ee
In Fig.~\ref{fig_nk_ueg}
we show $n_k$ as obtained from Monte Carlo simulations.

If we reproduce $n_k$ by Eq.~\eqref{eq_nk_twopole}
we obtain constraints for $a_{k,j}$ and $E_{k,j}$
which themselves put limits on the permissible values
of $M^{(1)}$, $M^{(2+)}$, and $M^{(3+)}$.
Since we may estimate $M^{(1)}$ and $M^{(2+)}$
according to Sec.~\ref{sec_ueg_first_two}
this means that reproducing $n_k$ with given
$M^{(1)}$ and $M^{(2+)}$ determines $M^{(3+)}$.
For $k<k_{\rm F}$ we have $n_k<1$ (see Fig.~\ref{fig_nk_ueg}).
The only way to realize this by Eq.~\eqref{eq_nk_twopole}
at zero temperature is $E_{k,1}<E_{\rm F}$, $E_{k,2}>E_{\rm F}$
and $a_{k,1}=n_{k}$, where $E_{\rm F}$ is the Fermi energy of
the interacting UEG, i.e., $E_{\rm F}\ne \frac{\hbar^2 k_{\rm F}^2}{2m}$.
According to Fig.~\ref{fig_nk_ueg} at
$k<k_{\rm F}$ we have $a_{k,1}>a_{k,2}$, because $a_{k,2}=1-a_{k,1}$.
Similarly, we find for $k>k_{\rm F}$:
$E_{k,1}<E_{\rm F}$, $E_{k,2}>E_{\rm F}$, $a_{k,1}=n_{k}$, $a_{k,1}<a_{k,2}$.
The jump of $n_{k}$ at $k=k_{\rm F}$ implies therefore
that $a_{k,1}=n_{k}$ and $a_{k,2}=1-a_{k,1}$ jump as well.
Consequently, $M_{k}^{(3+)}$ is predicted to jump by the two-pole
approximation, because $M_{k}^{(1)}$ and $M_{k}^{(2+)}$ do not jump.
However, only in the two-pole approximation this jump is unavoidable:
It can be avoided when more poles are used.

Consider for example the three pole approximation:
\bege\label{eq_spec_fun_ueg_threepole}
\frac{S_{k}(E-\mu)}{\hbar}=\sum_{j=1}^3 a_{k,j}\delta(E-E_{k,j}),
\ee
where we assume $E_{k,1}<E_{k,2}<E_{k,3}$.
This model predicts the momentum distribution to be
\bege
n_k=\sum_{j=1}^3 f(E_{k,j})a_{k,j}.
\ee
With this model we can reproduce Fig.~\ref{fig_nk_ueg}
at zero temperature by assuming that
$E_{k,1}<E_{\rm F}$, $E_{k,3}>E_{\rm F}$, and $E_{k_{\rm F},2}=E_{\rm F}$, i.e.,
only $E_{k,2}$ crosses the Fermi level at $k=k_{\rm F}$.
Consequently, setting $a_{k,1}+a_{k,2}=n_{k}$ for $k<k_{\rm F}$
and $a_{k,1}=n_{k}$ for $k>k_{\rm F}$ avoids the jump in $a_{k,j}$ if
we set additionally $a_{k_{{\rm F}-},1}=a_{k_{{\rm F}+},1}=n_{k_{{\rm F}+}}$, where
$k_{{\rm F}-}$ is infinitesimally below $k_{\rm F}$, while $k_{{\rm F}+}$ is
infinitesimally above $k_{\rm F}$.

The first-order exchange energy used in LDA corresponds to
the value of $\Sigma^{(0)}_{k}$ at $k=k_{\rm F}$ (see Eq.~\eqref{eq_vx_lda}
and Ref.~\cite{PhysRev.140.A1133}).
The jump of $a_{k,j}$ in the two-pole approximation is
a disadvantage when we try to construct a local potential
for $M^{(3+)}$ for MFbSDFT: Ideally we would like to include
information from both $k_{{\rm F}-}$ and  $k_{{\rm F}+}$ but
we can only use $k_{{\rm F}-}$, because it is not practical
to reproduce a jump with a local potential. While we 
will stick to the two-pole approximation in most of this work,
we present in Sec.~\ref{sec_algo_general} a generalization
of the algorithm described in Ref.~\cite{momentis}
for the construction of the spectral function from the first $2P$ moments.
This is a first step towards using more poles and
towards accounting for
the jump of $n_{k}$ at $k_{\rm F}$.

Ignoring the jump of $n_{k}$ at $k_{\rm F}$
we may compute $M^{(3+)}$ for given $M^{(2+)}$
by requiring $a_{k_{{\rm F}-},1}=n_{k_{{\rm F}-}}$ and
solving Eq.~\eqref{eq_ak1_analy}
for $M^{(3)}$.
The solution may be given in analytical form:
\bege\label{eq_v3+_compute}
\mathcal{V}^{(3+)}=-\left[
M^{(1)}
\right]^3+\frac{-B+Z\sqrt{B^2-4AC}}{2A},
\ee
where
\bege
A=1-4Q,
\ee
\bege
B=2X-4YQ,
\ee
\bege
C=X^2-4QK,
\ee
\bege\label{eq_Q}
Q=\left[
  n_{k_{{\rm F}-}}-\frac{1}{2}
\right]^2,
\ee
\bege
Y=4
\left[
M^{(1)}
\right]^3
-6M^{(1)}M^{(2)},
\ee
\bege
K=-3\left[
M^{(1)}
\right]^2
\left[
M^{(2)}
\right]^2+4\left[
M^{(2)}
\right]^3,
\ee
\bege
X=2\left[
M^{(1)}
\right]^3-
3\left[
M^{(1)}
\right]\left[
M^{(2)}
\right].
\ee
Eq.~\eqref{eq_ak1_analy}
has two solutions and
Eq.~\eqref{eq_v3+_compute}
yields the first solution when we set $Z=1$,
and the second solution when we set $Z=-1$.
These two solutions are not equivalent.
When we use the parameter-free model $\mathcal{V}^{(2+)}_{\rm F}$,
only $Z=1$ leads to significant improvement of the spectra.
It is therefore likely that another constraint, in
addition to the constraint of the momentum distribution, 
will select the appropriate sign of $Z$:
We suspect that the appropriate sign of $Z$ may be derived
by considering for example the additional constraint imposed
by Eq.~\eqref{eq_corr_from_spec} below.
However, in this work we choose the sign empirically,
and set $Z=1$ because it improves the spectra with the
parameter-free model $\mathcal{V}^{(2+)}_{\rm F}$.
However, when parameterized models for $\mathcal{V}^{(2+)}$
are used (see Sec.~\ref{sec_constru_pot} for a discussion of possible
parameterizations), one may obtain improved spectra
also with $Z=-1$, because one can find parameters for which
the computed spectra match experiment.

\begin{figure}
\includegraphics[angle=0,width=\linewidth]{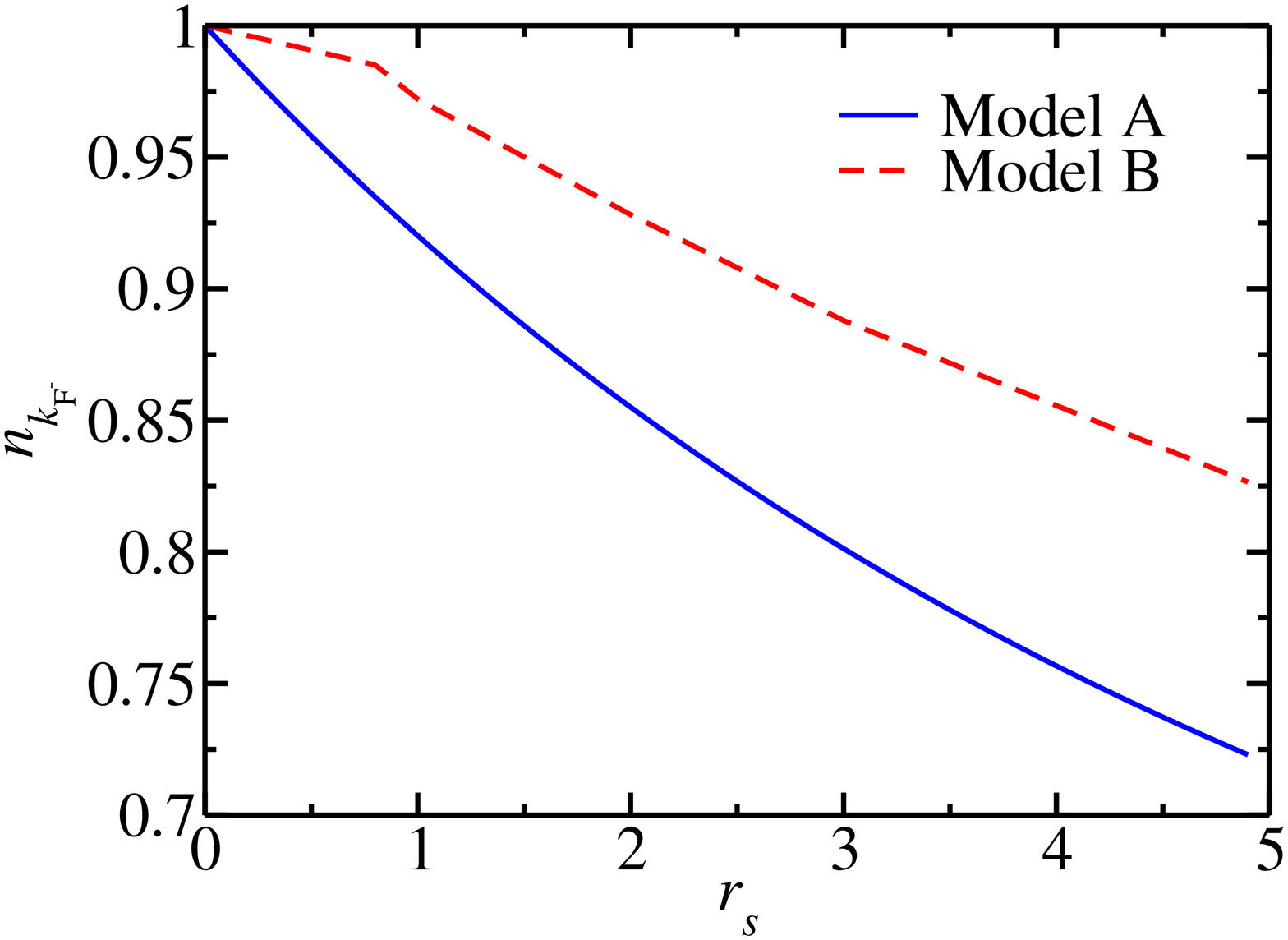}
\caption{\label{fig_kulik_compare}
  Plot of $n_{k_{\rm F}- }$ vs.\ $r_s$ for two different models of
  the momentum distribution. Model B: Ref.~\cite{PhysRevB.56.9970}.
  Model A: Ref.~\cite{PhysRevB.66.235116}. 
}
\end{figure}

Different models for $n_{k_{{\rm F}-}}$ are available.
One model is given by the direct fit of Monte Carlo
data~\cite{PhysRevB.56.9970}.
A second model additionally makes use of analytical results of the UEG~\cite{PhysRevB.66.235116}.
The behaviour of $n_k$ as
$k\rightarrow k_{\rm F}$ is quantitatively different between these
two models. We illustrate this in Fig.~\ref{fig_kulik_compare}.
A detailed comparison between the jumps of $n_{k}$ at $k_{\rm F}$ obtained by
various methods is
provided in Table~II of Ref.~\cite{PhysRevLett.107.110402}.
Moreover, the number of states in the interval $dk$ is
$n_{k}k^2 dk$. For the Hartree Fock model the maximum
of this function is always $k_{\rm F}$. However, for the refined
models of $n_k$ it may differ from $k_{\rm F}$. Therefore, the question arises
of whether one should use $n_{k_{{\rm F}-}}$
in Eq.~\eqref{eq_Q} or if one should replace this by the $k$ which
maximizes $n_{k}k^2$. In order to find out first how sensitive the
results are to the precise model of $n_{k_{{\rm F}-}}$ we tested \textit{model A}
and \textit{model B} (see Fig.~\ref{fig_kulik_compare}). We found that the results
depend strongly on the model for $n_{k_{{\rm F}-}}$ and that \textit{model B}
gives better results in Na and SrVO$_3$ when used together with $\mathcal{V}^{(2+)}_{\rm F}$.
Since the results depend strongly on the model of $n_{k_{{\rm F}-}}$
we expect that the construction of the third moment based on the momentum distribution
of the UEG can be improved further by improving the models of $n_{k_{{\rm F}-}}$
and perhaps also by using instead of $k_{\rm F}$ the $k$ that maximizes the function $n_{k}k^2$. 

Ref.~\cite{PhysRevB.69.045113} argues that it is likely that
the third moment is zero in the UEG. However, with the third moment
set to zero Eq.~\eqref{eq_ak1_analy} is generally not satisfied.
Therefore, with $M^{(3)}=0$ one cannot reproduce the momentum distribution
of the UEG at $k_{{\rm F}-}$ within the two-pole approximation. We suspect
that this remains true even when more poles are used. This suggests that
it is unlikely that the third moment is zero in the UEG.

\subsection{Construction of the moment potential}
\label{sec_constru_pot}
In this work we always construct $\mathcal{V}^{(3+)}$
according to Eq.~\eqref{eq_v3+_compute}.
Therefore, we do not need to develop a parameterized model
for $\mathcal{V}^{(3+)}$.
However, according to the discussion in Sec.~\ref{sec_ueg_first_two}
several plausible choices for  $\mathcal{V}^{(2+)}$ should be tested
and compared. The most obvious choice is
the parameter-free model $\mathcal{V}^{(2+)}_{\rm F}$
(see Eq.~\eqref{eq_vf2+} and Tab.~\ref{tab_compare_ueg}).

In Ref.~\cite{momentis} we have shown that spectral features
may often be reproduced with the model
\bege\label{eq_secmom_vc}
\mathcal{V}^{(2+)}_{\sigma}(\vn{r})=d^{(2+)}_{\sigma}[V_{c}(r_s)]^2,
\ee
where $d^{(2+)}_{\sigma}$ is a free parameter, which is chosen so that
the computed spectrum matches the experimental one as well as possible.

Since the model Eq.~\eqref{eq_secmom_vc} has only one free-parameter
the question arises how we can introduce additional parameters into it
to make it more flexible, while
keeping several of its key properties.
One key property is that it rises slower than $r_s^2$ when the density increases.
We suspect that this is the main reason why we found it easy to use in Ref.~\cite{momentis}.
A more flexible model with this key property is
\bege\label{eq_tanh_model}
\mathcal{V}^{(2+)}=\delta\frac{[\tanh(\beta r_s)]^{\gamma}}{r_s^2}.
\ee
A priori we suspect that this is a convenient model potential for inverse MFbSDFT,
because it has only three free parameters.

\section{Self-consistent computation of correlation functions from the spectral function}
\label{sec_selfcons_corr}
Ref.~\cite{PhysRevB.69.045113}
estimates the higher-order correlation
function $\left\langle
  c_{\vn{k}-\vn{q}'\sigma}^{\dagger}
  \rho_{\vn{q}}
  c_{\vn{k}-\vn{q}-\vn{q}'\sigma}
  \right\rangle$, which contributes to the second moment of the UEG,
  using the single Slater Determinant approximation, because Monte Carlo
  simulations have not yet been reported for this correlation function.
  Since the accuracy of using this approximation for this correlation function is not known, 
the accuracy of $\mathcal{V}_{\rm F}^{(2+)}$ and $\mathcal{V}_{\rm A}^{(2+)}$
given in Table~\ref{tab_compare_ueg}
is unclear as well.

In the single-band Hubbard model it is possible to express the most important
higher-order correlation functions that are needed for the two-pole approximation
in terms of the single-particle spectral function~\cite{book_Nolting}.
In the following we show that this is also possible for the
correlation
function $\left\langle
  c_{\vn{k}-\vn{q}'\sigma}^{\dagger}
  \rho_{\vn{q}}
  c_{\vn{k}-\vn{q}-\vn{q}'\sigma}
  \right\rangle$
  in the UEG.
  The Hamiltonian of the UEG is given by~\cite{PhysRevB.69.045113}
  \bege
H=\sum_{\vn{k}\sigma}\epsilon(\vn{k})c^{\dagger}_{\vn{k}\sigma}c_{\vn{k}\sigma}+\frac{1}{2}\mathcal{U},
  \ee
  where
\label{sec_selfcons_corr}
\bege
\mathcal{U}=\frac{1}{V}\sum_{\vn{q}}v(\vn{q})\left[\rho_{\vn{q}}\rho_{-\vn{q}}-\sum_{\vn{k}}c_{\vn{k}\sigma}^{\dagger}
  c_{\vn{k}\sigma}\right]
\ee
and $v(\vn{q})$ is the Coulomb potential. Using
\bege
    [\mathcal{U},c_{\vn{q}\sigma}^{\dagger}]_{-}
    =\frac{2}{V}\sum_{\vn{q}'}v(\vn{q}')c_{\vn{q}+\vn{q}'\sigma}^{\dagger}
    \rho_{\vn{q}'}
\ee
we may write
\bege\label{eq_cdag_rho_c}
\begin{aligned}
  &\frac{2}{V}\sum_{\vn{q}}v(\vn{q})
\langle
c_{\vn{k}-\vn{q}-\vn{q}'\sigma}^{\dagger}
\rho_{-\vn{q}}
c_{\vn{k}-\vn{q}'\sigma}
\rangle
=\\
=&\frac{2}{V}\sum_{\vn{q}}v(\vn{q})
\langle
c_{\vn{k}+\vn{q}-\vn{q}'\sigma}^{\dagger}
\rho_{\vn{q}}
c_{\vn{k}-\vn{q}'\sigma}
\rangle=\\
=&\langle
[\mathcal{U},c_{\vn{k}-\vn{q}'\sigma}]_{-}c_{\vn{k}-\vn{q}'\sigma}
\rangle.
\end{aligned}  
\ee
Our goal is to evaluate Eq.~\eqref{eq_cdag_rho_c}
with the help of the identity
\bege
\begin{aligned}
&\langle
    [H,c_{\vn{k}-\vn{q}'\sigma}]_{-}c_{\vn{k}-\vn{q}'\sigma}
\rangle
=\\
&=
    \int dE f(E)
      E
    S_{\vn{k}-\vn{q}'\sigma}(E-\mu).
\end{aligned}
\ee
For this purpose we use
\bege
\begin{aligned}
  &\left\langle
      \left[
        \sum_{\vn{k}',\bar{\sigma}}
        \epsilon(\vn{k}')
        c^{\dagger}_{\vn{k}'\bar{\sigma}}
        c_{\vn{k}'\bar{\sigma}},
c^{\dagger}_{\vn{k}-\vn{q}'\sigma}
      \right]_{-}
      c_{\vn{k}-\vn{q}'\sigma}\right\rangle\\
      &=\epsilon(\vn{k}-\vn{q}')
      \left\langle
      c^{\dagger}_{\vn{k}-\vn{q}'\sigma}
      c_{\vn{k}-\vn{q}'\sigma}\right\rangle \\
      &=      \epsilon(\vn{k}-\vn{q}')
      \int
      d E
      f(E)
      S_{\vn{k}-\vn{q}'\sigma}(E-\mu)
\end{aligned}
\ee
to obtain
\bege
\begin{aligned}
 &\frac{1}{V}\sum_{\vn{q}}v(\vn{q})
\langle
c_{\vn{k}-\vn{q}-\vn{q}'\sigma}^{\dagger}
\rho_{-\vn{q}}
c_{\vn{k}-\vn{q}'\sigma}
\rangle
=\\
&=\int
dEf(E)
S_{\vn{k}-\vn{q}'\sigma}(E-\mu)
\left[
E-\epsilon(\vn{k}-\vn{q}')
\right].
\\
 \end{aligned}
\ee
Consequently,
\bege\label{eq_corr_from_spec}
\begin{aligned}
  &-\frac{2}{V^2}\sum_{\vn{q}\vn{q}'}
  v(\vn{q})v(\vn{q}')
  \left\langle
  c_{\vn{k}-\vn{q}'\sigma}^{\dagger}
  \rho_{\vn{q}}
  c_{\vn{k}-\vn{q}-\vn{q}'\sigma}
  \right\rangle=\\
=&-\frac{2}{V^2}\sum_{\vn{q}\vn{q}'}
  v(\vn{q})v(\vn{q}')
  \left\langle
c^{\dagger}_{\vn{k}-\vn{q}-\vn{q}'\sigma}
    \rho_{-\vn{q}}
  c_{\vn{k}-\vn{q}'\sigma}  
  \right\rangle=\\
  =&-\frac{2}{V}
  \sum_{\vn{q}'}
  v(\vn{q}')
  \int
  dE
  f(E)
  S_{\vn{k}-\vn{q}'\sigma}(E-\mu)\times\\
&\times  
\left[
E-\epsilon(\vn{k}-\vn{q}')
\right].
\end{aligned}
\ee
The first line in Eq.~\eqref{eq_corr_from_spec}
is the contribution to the second moment of the UEG that
Ref.~\cite{PhysRevB.69.045113} has estimated based on the
single Slater Determinant approximation (see the last line
of Eq.~(22) in Ref.~\cite{PhysRevB.69.045113}). The last
line of Eq.~\eqref{eq_corr_from_spec} shows therefore that this
contribution may alternatively be computed from the spectral
function. If one approximates the spectral function of the UEG
by the two-pole approximation, one may therefore compute this
contribution directly and compare to the single Slater Determinant
approximation in order to check the validity of the latter.

\section{Construction of the spectral function from the first $\vn{2P}$ moments}
\label{sec_algo_general}
In Sec.~\ref{sec_ueg_twopole}
we have argued that meaningful approximations for the third moment
can be derived easily even when the third moment cannot be calculated directly from
existing Monte Carlo data on higher-order correlation functions of the UEG.
One may expect that a similar conclusion may be reached for even higher moments.
Of particular interest is the increase in precision with which the spectral
function of the UEG can be computed due to methodological advances
such as diagrammatic Monte Carlo~\cite{Haule2022_diagrammatic}
and coupled-cluster theory~\cite{PhysRevB.93.235139}.
If precise spectral functions of the UEG become available for a sufficient
set of density parameters $r_{s}$, one may compute the spectral moments, including
the higher ones, directly from them.
Since the precision of MFbSDFT is expected to increase with increasing number of
moments used, we demonstrate in the following that the spectral function
may be obtained generally from the first $2P$ ($P=1,2,3,4,5,\dots$) spectral
moment matrices by
diagonalizing a $PN\times PN$ matrix.

When we consider the first $2P$ spectral moment matrices of size $N\times N$,
we expect that the spectral function
may be approximated by~\cite{momentis}
\bege\label{eq_approxi_specfun}
\frac{S_{\sigma n m}(E)}{\hbar}=\sum_{p=1}^{P}\sum_{\gamma=1}^{N}
a_{\gamma p \sigma}
\mathcal{V}_{n\gamma p\sigma}\mathcal{V}^{*}_{m\gamma p\sigma}
\delta(E-E_{\gamma p\sigma}),
\ee
where $E_{\gamma p \sigma}$
are the $PN$ poles of the spectral function
for spin $\sigma$, $a_{\gamma p\sigma}$
are the corresponding spectral weights,
and $\vn{\mathcal{V}}_{\gamma p\sigma}$ are the corresponding
state vectors with components $\mathcal{V}_{n\gamma p\sigma}$.

In Ref.~\cite{momentis}
we have shown that for $P=2$ the spectral poles, spectral weights,
and state vectors may be obtained by diagonalizing a $2N\times 2N$ matrix.
In this case the $2N$ poles of the spectral function are simply the
$2N$ eigenvalues of the matrix $\vn{\mathcal{B}}^{(1)}$, which
is defined by
\bege\label{eq_bi_generali}
\vn{\mathcal{B}}^{(I)}=\left(
\begin{array}{cc}
\vn{M}^{(I)}
&\vn{B}_{I} \\
\vn{B}_{I}^{\dagger} &\vn{D}_{I}
\end{array}
\right),
\ee
where 
we determine the $\vn{B}_{I}$ and $\vn{D}_{I}$
so that
\bege\label{eq_bij}
\vn{\mathcal{B}}_{I+J}=\vn{\mathcal{B}}_{I}\vn{\mathcal{B}}_{J}
\ee
is satisfied.
In the case of $P=2$ the matrices $\vn{B}_{I}$
and $\vn{D}_{I}$ are $N\times N$ matrices and we restrict
$1\leq I \leq 2$, $1\leq J \leq 2$, and $2\leq I+J \leq 3$.

In the following we show that this approach may be
generalized to $P>2$.
When we take the first $2P$ moments,
we need $PN$ spectral poles. If these $PN$ spectral
poles are identical to the eigenvalues of a hermitian matrix,
this matrix has to be of the size $PN\times PN$.
We may stick to the matrices defined in
Eq.~\eqref{eq_bi_generali}, if we require
that
$\vn{B}_{I}$ be an $N\times (P-1)N$ matrix,
$D_{I}$ be a $(P-1)N\times (P-1)N$ matrix,
and $\vn{\mathcal{B}}^{(I)}$ be
a $PN\times PN$ matrix.
$\vn{M}^{(I)}$ is always a $N\times N$ matrix
irrespective of the choice of $P$.

Eq.~\eqref{eq_bij} may be written as four equations:
\bege\label{eq_firstof4}
\vn{M}^{(I+J)}=\vn{M}^{(I)}\vn{M}^{(J)}+\vn{B}_{I}\vn{B}_{J}^{\dagger},
\ee
\bege\label{eq_secondof4}
\vn{B}_{I+J}=\vn{M}^{(I)}\vn{B}_{J}+\vn{B}_{I}\vn{D}_{J},
\ee
\bege
\vn{B}^{\dagger}_{I+J}=\vn{B}^{\dagger}_{I}\vn{M}^{(J)}+\vn{D}_{I}\vn{B}_{J}^{\dagger},
\ee
and
\bege
\vn{D}_{I+J}=\vn{B}_{I}\vn{B}_{J}+\vn{D}_{I}\vn{D}_{J}.
\ee
When we set $I=J=1$ in Eq.~\eqref{eq_firstof4}
we obtain
\bege\label{eq_b1_ambi}
\vn{M}^{(2)}=\vn{M}^{(1)}\vn{M}^{(1)}+\vn{B}_{1}\vn{B}_{1}^{\dagger},
\ee
which we may use to determine $\vn{B}_{1}$ unambiguously (up to a phase factor)
when it is a square matrix, i.e., when $P=2$. However, $\vn{B}_{1}$ is not a square matrix  for $P>2$
and since it has as many rows as $\vn{M}^{(2)}$, but more
columns than rows, Eq.~\eqref{eq_b1_ambi} has no unique solution.
Therefore, we  build a square $(P-1)N\times (P-1)N$ matrix
$\mathscrbf{B}$ by combining the first $P-1$ matrices $\vn{B}_{1},\ldots \vn{B}_{P-1}$
as follows:
\bege
\label{eq_b_scrbf}
\mathscrbf{B}=\begin{pmatrix}
\vn{B}_{1}\\
\vdots\\
\vn{B}_{P-1}
\end{pmatrix}.
\ee
In order to find
an equation for $\mathscrbf{B}$
we combine the corresponding $(P-1)^2$ equations
of the type Eq.~\eqref{eq_firstof4}:
\begin{widetext}
\begin{equation}\label{eq_bbdag}
\mathscrbf{B}\mathscrbf{B}^{\dagger}=
\begin{pmatrix}
  \vn{M}^{(2)}-\vn{M}^{(1)}\vn{M}^{(1)} &\quad\vn{M}^{(3)}-\vn{M}^{(1)}\vn{M}^{(2)} &\quad\dots &\quad\vn{M}^{(P)}-\vn{M}^{(1)}\vn{M}^{(P-1)}\\
  \vdots                             &\vdots                             &\vdots &\vdots\\
  \vn{M}^{(P)}-\vn{M}^{(P-1)}\vn{M}^{(1)} &\quad\vn{M}^{(P+1)}-\vn{M}^{(P-1)}\vn{M}^{(2)} &\quad\dots &\quad\vn{M}^{(2P-2)}-\vn{M}^{(P-1)}\vn{M}^{(P-1)}
\end{pmatrix}
\end{equation}
\end{widetext}
Since $\mathscrbf{B}$ is a square matrix, Eq.~\eqref{eq_bbdag}
determines $\mathscrbf{B}$ unambiguously (up to a phase factor).
If the matrix on the right-hand side of Eq.~\eqref{eq_bbdag},
which we denote by $\mathscrbf{W}$,
is positive definite,
we obtain
\bege
\label{eq_b_scrbf_pos_def}
\mathscrbf{B}=\mathscrbf{U}\sqrt{\mathscrbf{D}},
\ee
where $\mathscrbf{D}$ is a diagonal matrix, which holds
the eigenvalues of $\mathscrbf{W}$ on its diagonals,
and $\mathscrbf{U}$ is the unitary transformation that
transforms $\mathscrbf{W}$ into the diagonal form:
\bege
\mathscrbf{W}=\mathscrbf{U}\mathscrbf{D}\mathscrbf{U}^{\dagger}.
\ee

From the equation
\bege\label{eq_d1_ambi}
\vn{B}_{2}=\vn{M}^{(1)}\vn{B}_{1}+\vn{B}_{1}\vn{D}_{1}
\ee
we may obtain $\vn{D}_{1}$ unambiguously when
$\vn{B}_{1}$, $\vn{B}_{2}$, and $\vn{M}^{(1)}$ are given
and when $P=2$. However, when $P>2$ Eq.~\eqref{eq_d1_ambi}
does not determine $\vn{D}_{1}$ uniquely,
because $\vn{B}_{1}$ has fewer rows than columns and the
corresponding system of coupled equations for $\vn{D}_{1}$ is
underdetermined. Moreover, we have not yet used the
moment $\vn{M}^{(2P-1)}$ in any of the
equations above. Thus, taking any particular solution
for $\vn{D}_1$, which satisfies Eq.~\eqref{eq_d1_ambi},
to set up the matrix $\vn{\mathcal{B}}^{(1)}$, cannot be
a correct approach, because no use is made of  $\vn{M}^{(2P-1)}$.
The consequence of this incorrect approach is that
Eq.~\eqref{eq_bij} is not satisfied for all permissible $I$ and $J$.
Therefore, we combine the first $P-1$ equations of the
type of Eq.~\eqref{eq_secondof4} into the equation
\bege\label{eq_d1}
\mathscrbf{B}\vn{D}_{1}=
\begin{pmatrix}
  \vn{B}_{2}-\vn{M}^{(1)}\vn{B}_{1}\\
  \vn{B}_{3}-\vn{M}^{(2)}\vn{B}_{1}\\
  \vdots\\
  \vn{B}_{P}-\vn{M}^{(P-1)}\vn{B}_{1}\\
\end{pmatrix}.
\ee
In this equation $\mathscrbf{B}$, and the moments
$\vn{M}^{(1)}, \vn{M}^{(2)}, \dots \vn{M}^{(P-1)}$ are known.
However, on the right hand side $\vn{B}_{P}$ occurs,
which we have not yet determined.

To determine $\vn{B}_{P}$
we combine the following $P-1$ equations
of the type Eq.~\eqref{eq_firstof4}
into one equation:
\bege\label{eq_bp}
\mathscrbf{B}\vn{B}^{\dagger}_{P}=
\begin{pmatrix}
  \vn{M}^{(P+1)}-\vn{M}^{(1)}\vn{M}^{(P)}\\
  \vn{M}^{(P+2)}-\vn{M}^{(2)}\vn{M}^{(P)}\\
  \vdots\\
  \vn{M}^{(2P-1)}-\vn{M}^{(P-1)}\vn{M}^{(P)}\\
\end{pmatrix}.  
\ee
We may solve this equation for $\vn{B}^{\dagger}_{P}$ and
subsequently we may solve Eq.~\eqref{eq_d1}
for $\vn{D}_{1}$.
Note that the highest moment, $\vn{M}^{(2P-1)}$,
is used in Eq.~\eqref{eq_bp}. 

Using $\vn{D}_{1}$ (from Eq.~\eqref{eq_d1}),
and $\vn{B}_{1}$ (the first $N$ rows
from $\mathscrbf{B}$, Eq.~\eqref{eq_b_scrbf}, which
can be computed according to Eq.~\eqref{eq_b_scrbf_pos_def})
we can construct the
matrix $\vn{\mathcal{B}}^{(1)}$
(Eq.~\eqref{eq_bi_generali} with $I=1$)
and 
diagonalize it:
\bege\label{eq_b1_ududag_general}
\vn{\mathcal{B}}^{(1)}=\vn{\mathcal{U}}\vn{\mathcal{D}}\vn{\mathcal{U}}^{\dagger}.
\ee
The normalized eigenvectors
of $\vn{\mathcal{B}}^{(1)}$
comprise the columns of the unitary matrix $\vn{\mathcal{U}}$.
We use
\bege\label{eq_spec_wei_general}
a_{j}=\sum_{i=1}^{N} \mathcal{U}_{i j} [\mathcal{U}_{i j}]^{*}.
\ee
to
compute the spectral weight of state $j$.
Since the index
$i$ runs only from 1 to $N$ in this summation
and not from 1 to $PN$,
$a_{j}$ may be smaller than one, e.g.,
when bands split into lower and
upper Hubbard bands.
We define the $N\times P N$
matrix $\vn{\mathcal{V}}$
by
\bege\label{eq_statevecs_general}
\mathcal{V}_{i j}=\frac{\mathcal{U}_{i j}}{\sqrt{a_{j}}},
\ee
where $i=1,...,N$,
and $j=1,...,PN$.
The spectral function is given by
\bege\label{eq_final_spec_general}
\frac{S_{ij}(E)}{\hbar}=\sum_{l=1}^{PN}a_{l}\mathcal{V}_{il}\mathcal{V}^{*}_{jl}\delta(E-E_{l}),
\ee
where $E_l=\mathcal{D}_{ll}$. By adding a spin index $\sigma$ and
an index $p$ to distinguish the different branches of the spectrum,
we may rewrite Eq.~\eqref{eq_final_spec_general}
in the form of Eq.~\eqref{eq_approxi_specfun}.

\section{Results}
\label{sec_results}
\subsection{Na}
\begin{figure}
\includegraphics[angle=0,width=\linewidth]{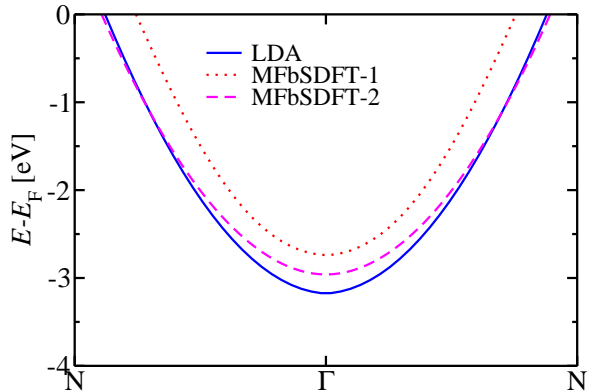}
\caption{\label{bandstruc_Na}
  Band structure of Na 
  as obtained from standard KS-DFT with LDA, and from MFbSDFT
  using Eq.~\eqref{eq_tanh_model} (MFbSDFT-1)
  and Eq.~\eqref{eq_vf2+} (MFbSDFT-2). $E_{\rm F}$ is the
  Fermi energy.
}
\end{figure}

The bandwidth of elemental Na is 3.3~eV in standard KS-DFT with LDA,
while it is only between 2.65~eV and 2.78~eV in
experiment~\cite{Mandal2022,PhysRevB.106.125138,PhysRevLett.60.1558}.
Here, following
Ref.~\cite{Mandal2022}, the bandwidth is defined as the distance of the
bottom of the conduction band to the Fermi level.
Using eDMFT the experimental bandwidth is reproduced
very well (2.84~eV in Ref.~\cite{Mandal2022}).
In Fig.~\ref{bandstruc_Na} we show that we may obtain
a similar reduction of the bandwidth within MFbSDFT:
The bottom of the conduction band is at 2.74~eV
in our calculation (called MFbSDFT-1 in the Figure).
To obtain this result we used Eq.~\eqref{eq_tanh_model}
with the parameters $\delta=1$~Ry$^2$, $\beta=0.5$, and $\gamma=2$.
This suggests that Eq.~\eqref{eq_tanh_model} is one
suitable model to approach the universal moment functional
in the spirit of a inverse MFbSDFT.

We also tested the parameter-free model $\mathcal{V}^{(2+)}_{\rm F}$
(see Eq.~\eqref{eq_vf2+} and Tab.~\ref{tab_compare_ueg}), which
yields a bandwidth of 2.96~eV (see MFbSDFT-2 in the Figure).
This is a remarkable improvement over
LDA, which suggests that it is likely that universal moment potentials
can be found for MFbSDFT. Additionally, this shows that one cannot
assess the effect of the moment potentials on the spectra by comparing only
$\mathcal{V}^{(2+)}$: Our recipe to compute $\mathcal{V}^{(3+)}$ from $\mathcal{V}^{(2+)}$
using the moment distribution of the UEG may lead to similar spectra even with
very different $\mathcal{V}^{(2+)}$, while setting e.g.\ $\mathcal{V}^{(3+)}=0$ will
typically lead to a strong sensitivity of the spectra on $\mathcal{V}^{(2+)}$.

There is an important qualitative difference between the calculations
MFbSDFT-1 and MFbSDFT-2: The MFbSDFT-1 dispersion is essentially given by a rigid shift
of the LDA dispersion. Consequently, charge neutrality implies that there have
to be states at lower energy that complete the 3s contribution to the valence charge.
Indeed there are such states in
our calculation.
However, different from the valence band satellite in Ni they are distributed over a broad
energy range. In contrast, in our MFbSDFT-2 calculation the average slope of the dispersion
is reduced combined with an upshift. Therefore, there is no need for states at lower energy
to satisfy charge neutrality. Indeed there are no states at lower energy in the MFbSDFT-2
calculation. Qualitatively, the MFbSDFT-2 calculation is in better agreement with
eDMFT, because eDMFT does not predict a rigid shift without any
change of slope~\cite{Mandal2022}. Moreover, the quantitative difference between eDMFT
and MFbSDFT-2 is quite small: 2.96~eV-2.84~eV=0.12~eV is the difference in the bandwidths
of these two calculations. Therefore, in Ni, the parameter-free $\mathcal{V}^{(2+)}_{\rm F}$
is almost as good as eDMFT as far as the prediction of the bandwidth is concerned.

Finally, we also tried Eq.~\eqref{eq_secmom_vc}. This model potential yields the bandwidth of 2.8~eV
when we use $d^{(2+)}=15.0$.

\subsection{SrVO$_3$}
\begin{figure}
\includegraphics[angle=0,width=\linewidth]{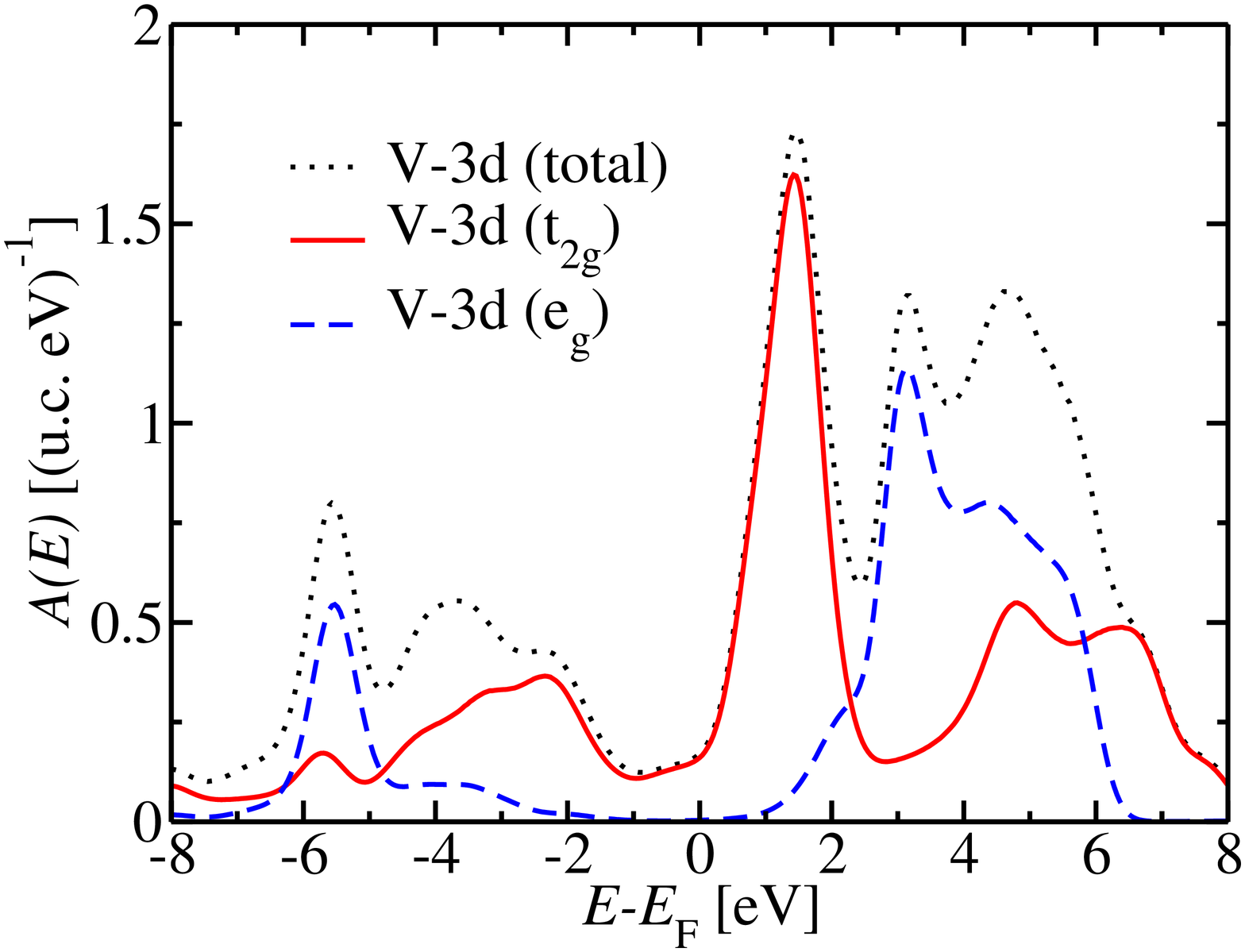}
\caption{\label{DOS_SrVO3}
  Orbitally decomposed spectral density $A(E)$ of SrVO$_3$ vs.\ energy $E$
  as obtained in MFbSDFT with
Eq.~\eqref{eq_secmom_vc}. $E_{\rm F}$ is the
  Fermi energy.
}
\end{figure}

According to experiment, the V-d-DOS in SrVO$_3$ is
distributed into three main
peaks, which are reproduced well in
LDA+DMFT~\cite{PhysRevLett.93.156402,PhysRevB.72.155106}.
In Ref.~\cite{momentis} we reproduced these spectral
features with MFbSDFT using Eq.~\eqref{eq_secmom_vc} with $d^{(2+)}$=100,
where we did not construct $\mathcal{V}^{(3+)}$ in order to satisfy the
momentum distribution of the UEG.
However, when we compute $\mathcal{V}^{(3+)}$ from Eq.~\eqref{eq_v3+_compute}
strong peaks at around -2~eV and 3~eV with the correct order of magnitude
of spectral density emerge already with a much smaller
$d^{(2+)}$=32 (using model A).
In Fig.~\ref{DOS_SrVO3} we show the orbitally decomposed
spectral density $A(E)$ of SrVO$_3$ obtained with $d^{(2+)}$=32.
Above the Fermi energy, our total V-d spectral density
agrees acceptably well with a LDA+DMFT calculation with a Hubbard $U$ of 6~eV (see Fig.~(8) in
Ref.~\cite{PhysRevB.77.205112}). However, our $A(E)$ exhibits more
V-d spectral weight in the region around -2~eV than Ref.~\cite{PhysRevB.77.205112}.
The distribution of the V-d spectral weight into the three regions
at around -2~eV, 1~eV, and 3~eV is in good agreement with experiment
and LDA+DMFT~\cite{PhysRevLett.93.156402,PhysRevB.72.155106}.

\begin{figure}
\includegraphics[angle=-90,width=\linewidth]{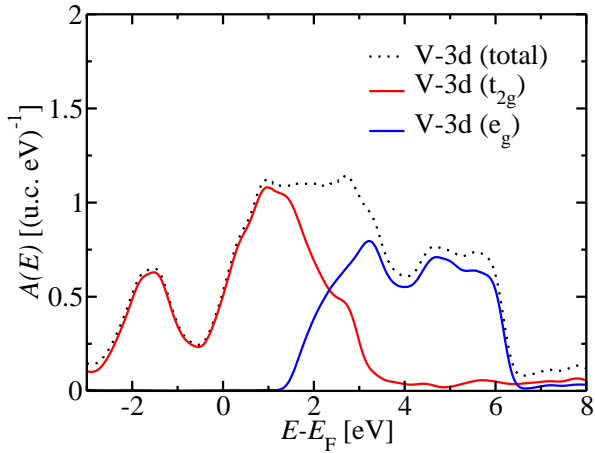}
\caption{\label{DOS_SrVO3_v2+f}
  Orbitally decomposed spectral density $A(E)$ of SrVO$_3$ vs.\ energy $E$
  as obtained in MFbSDFT with
Eq.~\eqref{eq_vf2+}. $E_{\rm F}$ is the
  Fermi energy.
}
\end{figure}

We also tested the parameter-free model $\mathcal{V}^{(2+)}_{\rm F}$
(see Eq.~\eqref{eq_vf2+} and Tab.~\ref{tab_compare_ueg}), which
yields $A(E)$ as shown in Fig.~\ref{DOS_SrVO3_v2+f} (using model B).
The agreement with experiment and LDA+DMFT~\cite{PhysRevLett.93.156402,PhysRevB.72.155106}
below the Fermi energy is remarkable.
Above the Fermi energy, the peaks at around 1~eV and 3~eV, which are
separate peaks in  experiment and LDA+DMFT~\cite{PhysRevLett.93.156402,PhysRevB.72.155106}
are merged together into a very broad double-peak.
Nevertheless, the overall redistribution of spectral weight constitutes
a remarkable improvement over
LDA, which corroborates the idea that universal moment potentials
should exist for MFbSDFT.

\subsection{Ni}
\begin{figure}
\includegraphics[angle=0,width=\linewidth]{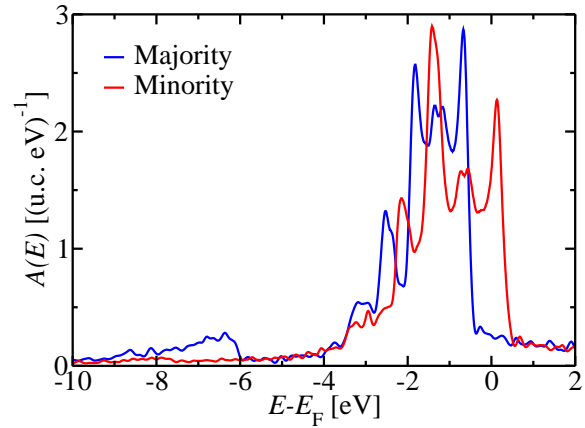}
\caption{\label{Ni_dos_vwn20}
  Spectral density $A(E)$ of Ni vs.\ energy $E$
  as obtained in MFbSDFT. $E_{\rm F}$ is the
  Fermi energy.
}
\end{figure}

In Ref.~\cite{momentis} we have computed the DOS of Ni
with $M^{(3+)}=0$, which yields an exchange splitting and
a bandwidth that agree with experiment. However,
with $M^{(3+)}=0$
we did not
obtain the strong spin-polarization of the satellite peak at
around 6~eV below the Fermi energy. Therefore, we discuss here
the spectral density
$A(E)$ of Ni obtained from MFbSDFT when $M^{(3+)}$ is constructed
according to the prescription given in Sec.~\ref{sec_ueg_twopole}
in order to show that this aspect can be improved.
We set $d^{(2+)}=20$ and first
compute $\mathcal{V}^{(2+)}$ and $\mathcal{V}^{(3+)}$ from the
charge density without considering the spin polarization.
In order to reproduce the magnetic moment, we need to spin-polarize
these potentials. We
use $\mathcal{V}^{(2+)}_{\sigma}=\zeta^6_{\sigma}\mathcal{V}^{(2+)}$
and $\mathcal{V}^{(3+)}_{\sigma}=\zeta^6_{\sigma}\mathcal{V}^{(3+)}$,
where $\zeta_{\sigma}=(1-\sigma (n_{\uparrow}-n_{\downarrow})/n)$.
With these parameters we obtain a magnetic moment of 0.6~$\mu_{\rm B}$.
In Fig.~\ref{Ni_dos_vwn20} we show the resulting $A(E)$   of Ni.
The valence band satellite at around 6~eV below the Fermi energy
is strongly spin-polarized in good agreement with experiment.
The width of the main band is reduced in comparison to LDA and therefore
it agrees better with experiment.
However, similar to standard KS-DFT with the PBE functional,
the exchange splitting is much larger than the experimental
value of 0.3~eV. Since our previous calculation in Ref.~\cite{momentis}
obtained the correct value of the exchange splitting, we suspect that the
reason for the overestimation of the exchange splitting in the present
calculation is the simplified treatment of the spin polarization.

\subsection{Pd}
Experimentally, a valence band satellite
has been found in Pd~\cite{exp_sat_pd}.
The inclusion of correlation effects reproduces
this satellite peak~\cite{PhysRevLett.45.482}.
When the correlations are treated with 
LDA+DMFT, additionally the equilibrium lattice
constant, the bulk modulus, and the band positions
are closer to experiment than in LDA~\cite{ele_struc_pd,lat_dyn_pd}.

\begin{figure}
\includegraphics[angle=0,width=\linewidth]{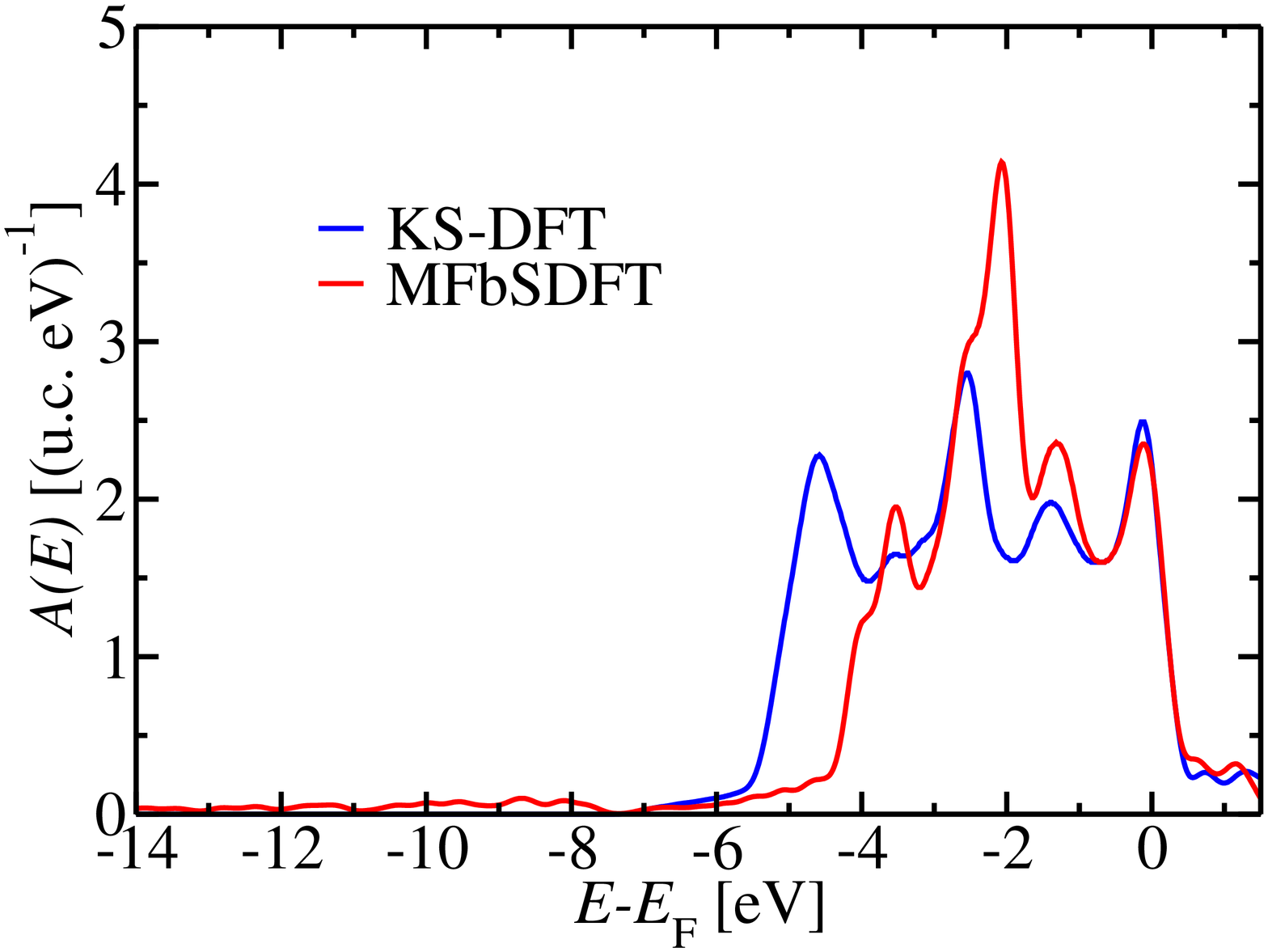}
\caption{\label{fig_DOS_Pd}
  Spectral density $A(E)$ of Pd vs.\ energy $E$
  as obtained in KS-DFT and in MFbSDFT. $E_{\rm F}$ is the
  Fermi energy.
}
\end{figure}

In Fig.~\ref{fig_DOS_Pd} we compare $A(E)$ calculated
within standard KS-DFT
using the PBE functional~\cite{PhysRevLett.77.3865}
and $A(E)$ obtained within
MFbSDFT.
The bandwidth of $A(E)$  in the region from
-6~eV up to the Fermi energy is slightly narrower
for the MFbSDFT calculation and several peaks 
are shifted to slightly higher energies similar to
LDA+DMFT calculations~\cite{ele_struc_pd}.
Moreover,
below -8~eV small satellite
peaks are visible in the MFbSDFT spectrum,
similar to the LDA+DMFT spectrum~\cite{ele_struc_pd}.
The spectral intensity in the
satellite region is comparable to the DMFT-result in the order of
magnitude~\cite{ele_struc_pd}.
These satellite peaks are missing in the KS-DFT spectrum.

To obtain these results we used  $d^{(2+)}=20$, i.e., we used essentially
the same potentials for the second and third moments as in the
MFbSDFT calculation of Ni.
The only difference is that we multiplied the $\mathcal{V}^{(3+)}$ as
obtained from Eq.~\eqref{eq_v3+_compute} by the factor 1.1 in order to
increase the intensity of the satellite peaks by roughly a factor of 2
such that they have the same intensity as in DMFT (without this factor of
1.1 the MFbSDFT spectra have satellite peaks with roughly half the intensity).
Remarkably, almost the same moment potentials
yield a smaller band-width reduction in Pd than in Ni, in agreement
with the experimental finding. Moreover, the satellite peaks in Pd
are much smaller than in Ni, also in agreement with experiment.

\section{Conclusions and open questions}
\label{sec_discussion}
While a model for the second moment of the UEG is available,
we found that it is often easier to use instead model potentials
for the second moment that rise less strong than $r_s^{-2}$ when
$r_s$ decreases. However, our parameterizations of the moment potentials
are comparable in the order of magnitude to the second moment of the UEG at small densities.

When we compute the third moment with the help
of the momentum distribution function of the UEG, the parameters in
our parameterizations of the moment potentials
vary by a factor of 2.1 between systems as different as
Na and SrVO$_3$.
While the hypothetical universality~\cite{momentis} of the moment potentials
implies that they are parameter free, a factor of 2.1 is not
much: This factor describes
the change of the second moment potential, which is quadratic in the
Coulomb interaction. In the Hubbard model, this contribution would
be quadratic in the Hubbard $U$ parameter and a change by a factor
of 2.1 would therefore correspond to a change of $U$ by a factor
$\sqrt{2.1}\approx 1.45$.
This small variation by a factor of 1.45 is also a big improvement
on our previous work Ref.~\cite{momentis}, where
we used $d^{(2+)}$ values as different as
15 and 100 in order to reproduce the experimental
spectra of Ni and SrVO$_3$, respectively. 
This larger difference arose because previously we chose the second
and third moments independently and we did not use the constraint of the
momentum distribution of the UEG.

Several reasons might explain why we have not found the
universal moment potentials yet:
First, it might be that with only 4 moments the hypothetical universal
moment potentials cannot reproduce the experimental spectra,
because more moments are necessary for this. When the number of moments
is not sufficient, it is plausible that a different choice of moment
potential is necessary for every system considered in order to optimize
the agreement with the experimental spectrum.
Second, gradient corrections might be necessary, because the
moment potentials are very steep at high density. In particular the
third moment potential is very inhomogeneous at high density, because
it rises much stronger than the second moment potential when $r_s$ becomes small.
Finally, we have not used many trial functions yet: If there is a single
parameter-free potential
which reproduces the experimental spectra of Ni, Na, Pd, and SrVO$_3$ 
it is probably not on the list of our model potentials yet.

Even if the universal moment potentials are not known yet,
one may alternatively construct element-specific moment potentials, i.e.,
build an \textit{empirical} variant of MFbSDFT.
The techniques necessary for empirical MFbSDFT are expected to be similar
to those employed in tight-binding quantum chemistry methods based on
density functional perturbation expansions~\cite{wcms.1493}.

\section{Summary}
\label{sec_summary}
We evaluate an existing model of the second moment of the UEG in order to
guide the search for suitable moment potentials in MFbSDFT.
Since models for the third moment of the UEG have not yet been developed
we use the known momentum distribution of the UEG in combination with the
two-pole approximation in order to compute the third moment of the UEG
when the second moment is given.
This leads to a parameter-free universal model for the second and third
moment potentials. These moment potentials correct the bandwidth
of Na towards the experimental value and also improve the spectrum of SrVO$_3$
significantly.
In Na the bandwidth correction is almost as good as in eDMFT.
The finding that the same parameter-free moment potentials improve the spectrum
of the very different systems Na and SrVO$_3$ corroborates the idea that universal
moment potentials exist for MFbSDFT.  

However, since the moment potentials
do not yet reproduce the experimental spectra perfectly,
we introduce parameterized model potentials for the second moment
in order to find out possible origins of the remaining discrepancies between
the theoretical and experimental spectra.
From the parameterized second moment, we
compute the third moment using the  momentum distribution of the UEG.
These techniques allow us to reproduce
the experimental spectra in Na, Ni, Pd, and SrVO$_3$ with various parameterizations
of the moment potentials. Importantly, the parameters in these parameterizations
vary only by about a factor of 2.1 between these very distinct systems.
While the yet unknown universal moment potentials are hypothetically parameter-free,
we suspect that the need for adjustable parameters in our model potentials
arises at least partly from using only the first four spectral moments. 
Therefore, we describe an efficient algorithm to
compute the spectral function from the first $2P$ moments.
This opens the perspective of using as many moments as necessary
to reproduce all spectral features accurately in MFbSDFT.
The higher moments may be constructed for example from
the spectral function of the UEG, which can be computed
with high precision using coupled-cluster theory
or diagrammatic Monte Carlo.

When the third moment is computed from the second moment using  the
momentum distribution of the UEG, it rises very strongly at high density.
Therefore, we expect that a second reason for the need for adjustable parameters in our model potentials is that we have not included gradient corrections.
\section*{Acknowledgments}
The project is funded by the Deutsche
Forschungsgemeinschaft (DFG, German Research Foundation) $-$ TRR 288 $-$ 422213477 (project B06),
CRC 1238, Control and Dynamics of Quantum Materials: Spin
orbit coupling, correlations, and topology (Project No. C01),
SPP 2137 ``Skyrmionics",  and Sino-German research project
DISTOMAT (DFG project MO \mbox{1731/10-1}).
We also
acknowledge financial support from the European Research
Council (ERC) under the European Union’s Horizon 2020
research and innovation program (Grant No. 856538, project
``3D MAGiC'') and computing resources granted by the Jülich Supercomputing
Centre under
project No. jiff40.

\bibliography{momdis}

\end{document}